\documentclass{article}
\pdfoutput=1
\usepackage{graphicx,xspace,url}
\usepackage{booktabs,color}
\usepackage{listings} 
\usepackage{trcover}
\newcommand {\omp}{OpenMP\xspace}

\newcommand{\code}[1]{\texttt{\small #1}\xspace}

\title{FastFlow: Efficient Parallel Streaming Applications on Multi-core}
\trnum{TR-09-12}
\author{Marco Aldinucci\thanks{Computer Science Department, University
    of Torino, Italy. Email: adinuc@di.unito.it} \and Massimo Torquati
  \and Massimiliano Meneghin}  
\date{September 2, 2009}

\begin{document}

\maketitle

\begin{abstract}
  Shared memory multiprocessors come back to popularity thanks to
  rapid spreading of commodity multi-core architectures.  As ever,
  shared memory programs are fairly easy to write and quite hard to
  optimise; providing multi-core programmers with optimising tools and
  programming frameworks is a nowadays challenge.  Few efforts have
  been done to  support effective streaming applications on these
  architectures. In this paper we introduce FastFlow, a low-level
  programming framework based on lock-free queues explicitly designed
  to support high-level languages for streaming applications.
  We compare FastFlow with state-of-the-art programming frameworks
  such as Cilk, OpenMP, and Intel TBB. We experimentally demonstrate 
  that FastFlow is always more efficient than all of them in a set of
  micro-benchmarks and on a real world application; the speedup edge of FastFlow 
  over other solutions might be bold for fine grain tasks, as an
  example   +35\% on OpenMP, +226\% on Cilk, +96\% on TBB for the
  alignment of protein P01111 against UniProt DB using Smith-Waterman algorithm.
\end{abstract}

\section{Introduction}
\label{sec:intro}

The recent trend to increase core count in commodity processors has led to
a renewed interest in the design of both methodologies and mechanisms
for the effective parallel programming of shared memory computer
architectures. Those methodologies are largely based on traditional
approaches of parallel programming. 

Typically, low-level approaches provides the programmers only with
primitives for flows-of-control management (creation, destruction),
their synchronisation and data sharing, which are usually accomplished
in critical regions accessed in mutual exclusion (mutex). As an
example, POSIX thread library can be used to this purpose. Programming parallel complex
applications is this way is certainly hard; tuning them for
performance is often even harder due to the non-trivial effects induced
by memory fences (used to implement mutex) on data replicated in
core's caches. 

Indeed, memory fences are one of the key sources of
performance degradation in communication intensive (e.g. streaming)
parallel applications. Avoiding memory fences means not only avoiding
locks but also avoiding any kind of atomic operation in memory
(e.g. Compare-And-Swap, Fetch-and-Add). While there exists several
assessed fence-free  solutions for \emph{asynchronous symmetric} communications\footnote{Single-Producer-Single-Consumer (SPSC) queues
\cite{Lamport}.}, these results cannot be easily extended to
\emph{asynchronous
  asymmetric} communications\footnote{Multiple-Producer-Multiple-Consumer queues
  (MPMC).}, which are necessary to support arbitrary streaming
networks.

A first way to ease programmer's task and improve program efficiency consist
in to raise the level of abstraction of concurrency management
primitives. As an example, threads might be abstracted out in
higher-level entities that can be pooled and scheduled in user space
possibly according  to specific strategies to minimise cache
flushing or maximise load balancing of cores. Synchronisation
primitives can be also abstracted out and associated to semantically 
meaningful points of the code, such as function calls and returns,
loops, etc. Intel \emph{Threading Building Block} (TBB)
\cite{intel:skeletons:tbb}, \emph{OpenMP} \cite{openMP}, and
\emph{Cilk} \cite{trasnmem:scicomp:05} all provide this 
kind of abstraction (even if each of them in its own way). 

This kind of abstraction significantly simplify the hand-coding of applications
but it is still too low-level to effectively automatise the
optimisation of the parallel code: here the major weakness lies in the
lack of information concerning the \emph{intent} of the code (idiom recognition
\cite{idiom:sc:95});  inter-procedural/component optimisation further
exacerbates the problem. The generative approach focuses on synthesising
implementations from higher-level specifications rather than
transforming them. From this approach, programmers' intent is captured by the
specification. In addition, technologies for code generation are
well developed (staging, partial evaluation, automatic programming,
generative programming). Both TBB and OpenMP follow this approach. The
programmer is required to explicitly define parallel behaviour by
using  proper constructs \cite{lithium:sem:CLSS}, which clearly delimit the interactions among
flows-of-control, the read-only data, the associativity of
accumulation operations, the concurrent access to shared data
structures. 

However, the  above-mentioned programming framework for  multi-core architectures are not
specifically designed to support streaming applications.  
The only pattern that fits this usage is TBB's \emph{pipeline} construct,
which can be  used to describe only a linear chain of filters; none of those
natively support any kind of \emph{task farming} on stream items
(despite it is a quite common pattern).
%depite  in spite of the 
%this skeleton, in either the stateless or stateful variants, is quite common 
%in streaming applications.

The objective of this paper is threefold: 

\begin{itemize}
\item To introduce FastFlow, i.e. low-level methodology supporting lock-free
  (fence-free) Multiple-Producer-Multiple-Consumer (MPMC)
  queues able to support low-overhead high-bandwidth multi-party
  communications in multi-core architectures, i.e. any \emph{streaming network},
  including cyclic graphs of threads.
\item To study the implementation of the farm streaming network using
  FastFlow \emph{and} the most popular programming frameworks for
  multi-core architectures (i.e. TBB, OpenMP, Cilk).
\item To show that FastFlow farm is generally faster than the other
  solutions on both a synthetic micro-benchmark and a real-world
  application, i.e. the Smith-Waterman local sequence alignment
  algorithm (SW). This latter comparison will be performed using
  the same ``sequential'' code in all implementations, i.e.  the x86/SSE2
  vectorised code derived from Farrar's high-performance
  implementation \cite{FarrarCell}. We will also show that the
  FastFlow implementation is faster than the state-of-the-art,
  hand-tuned parallel version of the Farrar's code (SWPS3 \cite{FarrarSSE2}).
\end{itemize}

In the longer term, we envision FastFlow as the part of a run-time
support of a set of high-level streaming skeletons for multi- and
many-core, either in insulation or as extension of the TBB programming
framework.

\section{Related Works}

The stream programming paradigm offers a promising approach for
programming multi-core systems. Stream languages are motivated by the
application style used in image processing, networking, and other
media processing domains.  Several languages and libraries are available
for programming stream applications, but many of them are oriented to
coarse grain computations. 
%Some are general purpose programming languages that hide the
%underlying architectural specificity.  
% Stream languages enable the explicit specification of
% producer-consumer parallelism between coarse grain units of
% computation; 
Example are \emph{StreamIt} \cite{streamIt}, \emph{Brook}
\cite{Brook}, and \emph{CUDA} \cite{CUDA}. Some other languages, as
TBB, provide explicit mechanisms for both streaming and other parallel
paradigm, while some others, as \emph{OpenMP} \cite{openMP} and 
\emph{Cilk} mainly offers  mechanisms for Data Parallelism and
Divide\&Conquer computations. These mechanisms can be also exploited to implement
streaming applications, as we shall show in Sec.~\ref{sec:farm}, but this requires
a greater programming effort with respect to the other cited languages.

StreamIt is an explicitly parallel programming language based on the
Synchronous Data Flow (SDF) programming model.  A StreamIt program is
represented as a set of autonomous actors %(called filters)
that communicate through first-in first-out (FIFO) data channels.
StreamIt contains syntactic constructs for defining programs
structured as task graphs, where each tasks contain Java-like
sequential code. The interconnection types provided by are: {\it
  Pipeline} for straight task combinations, {\it SplitJoin} for
nesting data parallelism and {\it FeedbackLoop} for connections from
consumers back to producers.  The communications are implemented
either as shared circular buffers or message passing for small amounts
of control information.

Brook \cite{Brook} provides extensions to C language with single
program multiple data (SPMD) operations that work on streams.
User defined functions operating on stream elements are called \emph{kernels} and 
can be executed in parallel.  
Brook kernels feature a blocking behaviour: the execution of a kernel
must complete before the next kernel can execute.  This is the same
execution model that is available on graphics processing units
(GPUs), which are indeed the main target of this programming
framework. In the same class can be enumerated CUDA \cite{CUDA}, which
is an infrastructure from NVIDIA. In addition, CUDA programmers are
required to use low-level mechanisms to explicitly manage the various level
of the memory hierarchy. 

Streaming applications are also targeted by TBB
\cite{intel:skeletons:tbb} through the \emph{pipeline}
construct. FastFlow -- as intent --  is methodologically similar to TBB, since it aims
to provide a library of explicitly parallel constructs
(a.k.a. parallel programming paradigms or skeletons) that extends the
base language (e.g. C, C++, Java). However, TBB does not
support any kind of non-linear streaming network, which therefore has
to be embedded in a pipeline. This has a non-trivial programming and
performance drawbacks since pipeline stages should bypass data that
are not interested with.
 
\emph{OpenMP} \cite{openMP} and \emph{Cilk} \cite{BlumofeJoKu96} are
other two very popular thread-based frameworks for multi-core
architectures (a in deep language descriptions is reported in
\ref{openmp} and \ref{cilk} sections). \emph{OpenMP} and \emph{Cilk}
mostly target Data Parallel and Divide\&Conquer programming paradigms, 
respectively.  OpenMP for example has only recently extended (3.0
version) with a {\it task} construct to manage the execution of a set
of independent tasks. The fact that the two
languages do not provide first class mechanisms for streaming
applications is reflected in their characteristic of well performing
only with coarse- and medium-grained computations, as we see in Sec.~\ref{sec:exp}.

At the level of communication and synchronisation mechanisms, 
Giacomini et al. \cite{fastforward:ppopp:08} highlight that
traditional locking queues feature a high overhead on today
multi-core.  Revisiting Lamport work \cite{Lamport}, which proves the
correctness of wait-free mechanisms for concurrent
Single-Producer-Single-Consumer (SPSC) queues on system with memory
sequential consistency commitment, they proposed a set of wait-free
and cache-optimised protocols. They  also prove the
performance benefit of those mechanisms on pipeline applications on
top of today multi-core architectures.  Wait-free protocols are a
subclass of lock-free protocols exhibiting even stronger properties: roughly
speaking lock-free protocols are based on retries while wait-free protocols
guarantee termination in a finite number of steps.

Along with SPSC queues, also MPMC queues are required to provide a
complete support for streaming 
networks. Those kind of data structures represent a more general
problem than SPSC one, and various works has been presented in
literature \cite{mpmc1,mpmc2,mpmc3,mpmc4}. Thanks to the structure of
streaming applications, we avoid the problem of managing directly MPMC
queue: we exploit multiple SPSC queues to implement MPSC, SCMP and
MPMC ones.

Therefore exploiting a wait-free SPSC also for implementing more
complex shared queues, FastFlow widely extend the work of Giacomini et
al., from simple pipelines to {\it any streaming networks}. We show
effective benefits of our approach with respect to the other languages
TBB, OpenMP and Cilk.

\section{Stream Parallel Paradigm: the Farm Case}
\label{sec:farm}

Traditionally types of parallelisms are categorised in three main
classes:

\begin{itemize}
\item \emph{Task Parallelism}. Parallelism is explicit in the
  algorithm and consists of running the same or different code on
  different executors (cores, processors, etc.). Different
  flows-of-control (threads, processes, etc.)  communicate with one
  another as they work. Communication takes place usually to pass data
  from one thread to the next as part of a graph. 
\item \emph{Data Parallelism} is a method for parallelising a single task by
  processing independent data elements of this task in parallel. The
  flexibility of the technique relies upon stateless processing
  routines implying that the data elements must be fully
  independent. Data Parallelism also support \emph{Loop-level
    Parallelism} where successive iterations of a loop working on independent or
  read-only data are parallelised in different flows-of-control
  (according to the model \emph{co-begin/co-end}) and  concurrently executed. 
\item \emph{Stream Parallelism}  is method for parallelising the
  execution (a.k.a. filtering) of a stream of tasks by segmenting the
  task into a series of \emph{sequential}\footnote{In the case of
    total sequential stages, the method is also known as
    \emph{Pipeline Parallelism}.} or \emph{parallel}
  stages. This  method can be also applied when there exists a
  \emph{total} or \emph{partial} order, respectively, in a computation preventing the
  use of data or task parallelism.  This might also come from the
  successive availability of input data along time (e.g. data flowing
  from a device). By processing data elements in order, local state
  may be either maintained 
 in each stage or distributed (replicated, scattered, etc.)
 along streams. Parallelism is achieved by running each stage
 simultaneously on \emph{subsequent} or \emph{independent} data elements.  
\end{itemize}

These basic form of parallelism are often encoded in high-level
paradigms (a.k.a. \emph{skeletons}) to be encoded in programming language
construct.
Many skeletons appeared in literature in the last two decades covering
many different usage schema of the three classes of  parallelism, on
top of both the message 
passing \cite{ske:homepage, darlington:parle:93,
  serot:taggedtoken:ppl:2001, assist:imp:europar:03,  lithium:fgcs:03,
  DBLP:journals/ppl/BischofGL03, kuchen-farm,
  beske:pdp:09,mapreduce:google:04,assist:cunhabook:05} and shared memory 
\cite{eskimo:PPL:03,intel:skeletons:tbb} programming models. 

 As an example, the farm skeleton models the functional replication and consists of running 
multiple independent tasks in parallel or filtering many successive
tasks of a stream in parallel. It typically consists of
two main entities: a master (or scheduler) and multiple workers (farm
is also known as \emph{Master-Workers}).
The scheduler is responsible for distributing the input task (in case by 
decomposing the input task into small tasks) toward the worker pool, as well 
as for gathering the partial results in order to produce the final result 
of the computation. 
The worker entity get the input task, process the task, and send the result 
back to the scheduler entity. Usually, in order to have pipeline
parallelism between the scheduling phase and the gathering phase, the
master entity is split in two main entities: respectively the
\emph{Emitter} and the \emph{Collector}. Farm can be declined in many
variants, as for example with stateless workers, stateful workers
(local or shared state, read-only or read/write), etc.

The farm skeleton is quite useful since it can exploited in many
streaming applications. In particular, it can be used in any
\emph{pipeline} to boost the service time of slow 
stages, then to boost the whole pipeline  \cite{pdcs:nf:99}.

As mentioned in previous section, several programming framework
for multi-core offer Data Parallel and Task Parallel skeletons, only few of them
offer Stream Parallel skeletons (such as TBB's pipeline), none of them
offers the farm. In the following we study the implementation of the
farm for multi-core architectures. In Sec.~\ref{sec:farm:fastflow} we 
introduce a very efficient implementation of the farm construct in FastFlow,
and we propose our implementation using other well-known frameworks such 
as OpenMP, Cilk, and TBB. The performance are compared in Sec.~\ref{sec:exp}.

\subsection{FastFlow Farm}
\label{sec:farm:fastflow}
FastFlow aims to provide a set of low-level mechanisms able  to support
low-latency and high-bandwidth data flows in a network of threads
running on a SCM. These flows, as typical in streaming
applications, are supposed to be mostly unidirectional and
asynchronous. On these architectures, the key issues regard memory
fences, which are required to keep the various caches coherent. 

FastFlow currently provides the programmer with two basic
mechanisms: MPMC queues and a memory allocator. The memory allocator,
is actually build on top of MPMC queues and can be substituted either 
with OS standard allocator (paying a performance penalty) or a third-party 
allocator (e.g. Intel TBB scalable allocator \cite{intel:skeletons:tbb}).

The key intuition underneath FastFlow is to provide the programmer with
lock-free MP queues and MC queues (that can be used in pipeline to build
MPMC queues) to support fast streaming networks. 
Traditionally, MPMC queues are build as passive entities: threads
concurrently synchronise (according to some protocol) to access data;
these synchronisations are usually supported by one or more atomic
operations (e.g. Compare-And-Swap) that behave as memory fences.  
FastFlow design follows a different approach: in order to avoid any memory
fence, the synchronisations among queue readers or writers are
arbitrated by an active entity (e.g. a thread), as shown in
Fig.~\ref{fig:FF:MDF}. We call these entities \emph{Emitter} (E) or
\emph{Collector} (C) according to their role; they actually read an
item from one or more lock-free SPSC queues and write onto one or more
lock-free SPSC queues. This requires a memory copy but no atomic
operations (this is a trivial corollary of lock-free SPSC correctness
\cite{fastforward:ppopp:08}). Notice that, FastFlow networks do not suffer from the ABA
problem \cite{ABA:98} since MPMC queues are build explicitly
linearising correct SPSC queues using Emitters  and Collectors.

The performance advantage of this solution descend from the higher 
speed of the copy with respect to the memory fence, that advantage 
is further increased by avoiding cache invalidation triggered by fences. 
This also depends on the size and the memory layout of copied data. 
The former point is addressed using data pointers instead of data, and 
enforcing that the data is not concurrently written: in many cases this can 
be derived by the semantics of the skeleton that has been implemented 
using MPMC queues (as an example this is guaranteed in a stateless
farm and many other cases). 

When using dynamically allocated memory, the memory allocator plays an
important role in term of performance. Dynamic memory allocators
(\code{malloc}/\code{free}) 
rely on mutual exclusion locks for protecting the consistency of their shared
data structures under multi-threading. Therefore, the use of memory allocator
may subtly reintroduce the locks in the lock-free application.
For this reason, we decided to use our own custom memory allocator,
which has specifically optimised for SPMC pattern.
The basic assumption is that, in streaming application, typically, one thread 
allocate memory and one or many other threads free memory. 
This assumption permits to develop a multi-threaded memory allocator that
use SPSC channels between the allocator thread and the generic thread that
performs the free, avoiding the use of costly lock based protocols for
maintaining the memory consistency of the internal structures.
Notice however, the FastFlow allocator is not a general purpose
allocator and it currently exhibits several limitations, such as a
sub-optimal space usage. The further development of FastFlow allocator
is among future works.

\subsubsection{Pseudo-code}
%\paragraph{Pseudocode}
\begin{figure*}
\begin{center}
%\vspace{-3ex}
%\includegraphics[scale=0.6]{mdf_interp_2}
\includegraphics[width=0.7\linewidth]{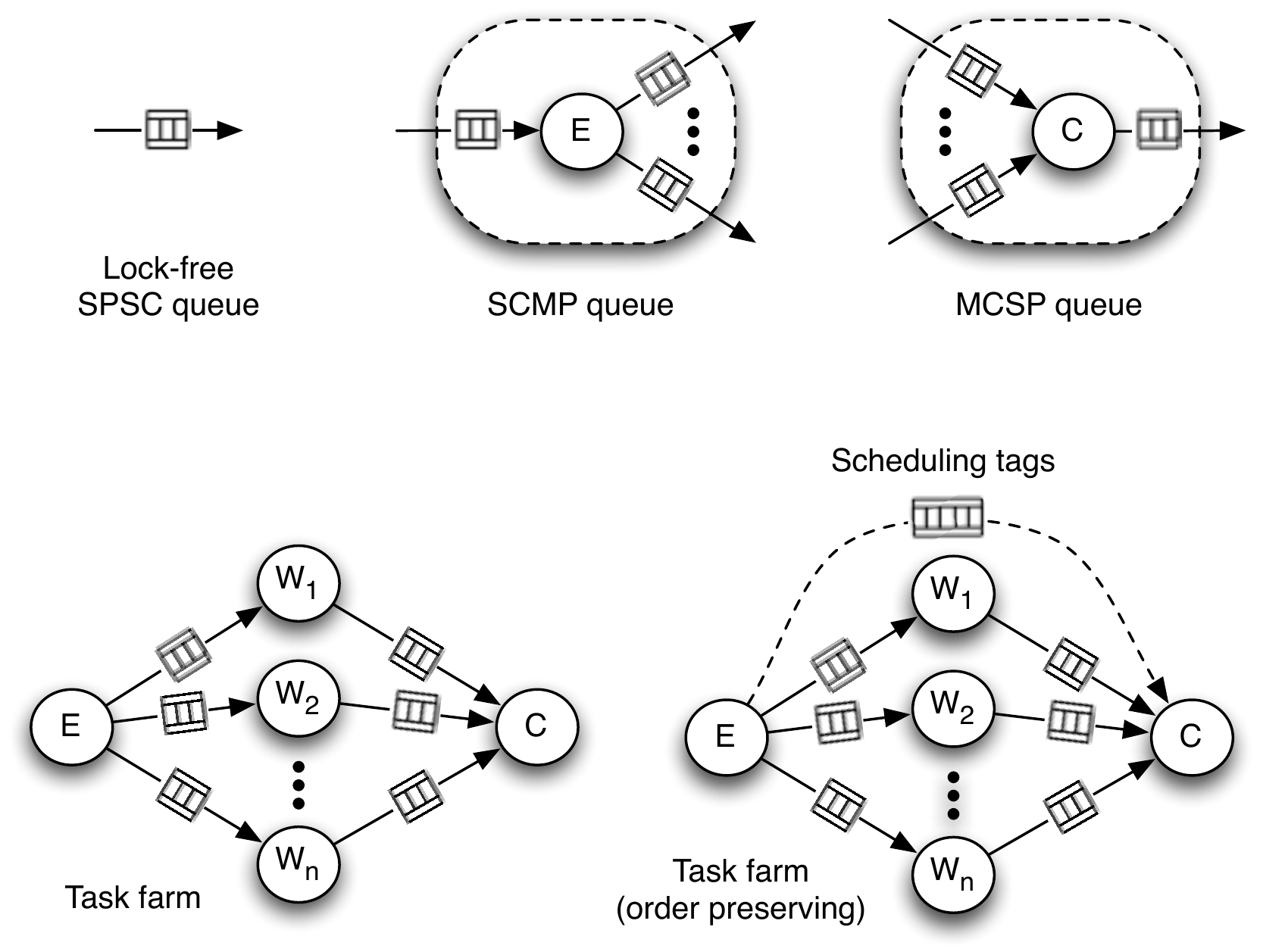}
\end{center}
%\vspace{-2ex}
\caption{FastFlow concepts: Lock-free SPSC queue, SPMC queue, MCSP queue,
  task farm, and order preserving task farm. \label{fig:FF:MDF}}
\end{figure*}
\lstnewenvironment{FastFlow}[2] 
{\lstset{ %
language=C++,                % choose the language of the code
basicstyle=\ttfamily\footnotesize,       % the size of the fonts that are used for the code
%numbers=left,                   % where to put the line-numbers
%numberstyle=\footnotesize,      % the size of the fonts that are used for the line-numbers
%stepnumber=1,                   % the step between two line-numbers. If it's 1 each line will be numbered
%numbersep=5pt,                  % how far the line-numbers are from the code
backgroundcolor=\color{white},  % choose the background color. You must add \usepackage{color}
showspaces=false,               % show spaces adding particular underscores
showstringspaces=false,         % underline spaces within strings
showtabs=false,                 % show tabs within strings adding particular underscores
%frame=single,                   % adds a frame around the code
frame=shadowbox
tabsize=2,                      % sets default tabsize to 2 spaces
captionpos=b,                   % sets the caption-position to bottom
breaklines=true,                % sets automatic line breaking
breakatwhitespace=false,        % sets if automatic breaks should only happen at whitespace
escapeinside={\%*}{*)},          % if you want to add a comment within your code
%MIEI =P
mathescape=true,
columns=flexible,
morekeywords={main, ff::, ff_TaskFarm, ff_node, add_emitter, add_collector, add_worker},
%backgroundcolor=\color{green},
%#1,
%label={code:#2}
}}{}
\begin{figure}[!h]
\begin{FastFlow}{}{}
class Emitter: public ff::ff_node {
public:
   void * svc(void *) {
     while($\exists\;newtask$){
       $newtask$ = create_task();
       return $newtask$;
     }
     return NULL; //EOS
   }
};

class Collector: public ff::ff_node {
public:
   void * svc(void * task) {
     collect_task($task$);
     return task;
   }
};

class Worker: public ff::ff_node {
public:
   void * svc(void * task) {
     compute_task(task);
     return task;
   }
};

int main (int argc, char *argv[]) {
  Emitter   E;
  Worker    W;
  Collector C;
  ff::ff_TaskFarm farm(nworkers);
  farm.add_emitter(E);
  farm.add_worker(W);
  farm.add_collector(C);

  farm.run();
}
\end{FastFlow}
\caption{Farm structure in FastFlow.}
\label{code:farm_ff}
\end{figure}

The structure of the farm paradigm in FastFlow is sketched in
Fig.~\ref{code:farm_ff}. 
The \code{ff\_TaskFarm} is a C++ class interface
that implements the parallel farm construct composed by an Emitter and
an optional Collector (also see Fig.~\ref{fig:FF:MDF}). The number of workers should be
fixed at the farm object creation time. 

 The usage of the interface is straightforward: Firstly,  Emitter, Worker
and Collector classes are defined deriving the \code{ff\_node}
class. Secondly, for each of the three classes the abstract method
\code{svc} (i.e. \emph{service} method) should be implemented. The
method contains the sequential code of the worker entity. Finally, 
the three objects are registered with the object of the class
\code{ff\_TaskFarm}.

In Fig.~\ref{code:farm_ff}, the Emitter produces a new
task each time the \code{svc} method is called. The run-time support is
in charge to schedule the tasks to one of the available workers. The
scheduling can be performed according to several policies, from simple
round-robin to an user-defined stateful policy \cite{assist:imp:europar:03}. 
Observe that ordering of tasks flowing through the farm, in general,
is not preserved. However, task ordering can be ensured either using
the same deterministic policy for task scheduling and collection, or
by dynamically tracking scheduling choices and performing the
collection accordingly. This latter solution, schematised in
Fig.~\ref{fig:FF:MDF} (order preserving task farm), is actually derived from tagged-token macro
data-flow architecture \cite{tagged-token:mit:90, lithium:fgcs:03,serot:taggedtoken:ppl:2001}.

\subsection{OpenMP Farm}
\label{openmp}
The \omp{} ({\it Open Multi-Processing}) \cite{openMP,openmpTask} is a set of standardised API
developed to extend C, C++ and Fortran sequential languages in order
to support shared memory multiprocessing programming. \omp{} is based on
compiler directives, library routines, and environment variables that
can be used to transform a sequential program into a thread-based
parallel program.

A  key concept is that a well written \omp{} program should result in a completely correct
sequential program when it is compiled without any \omp{}
supports. Therefore, all the \omp{} directives are implemented as
\emph{pragmas} into the target languages:
they are exploited by those compilers featuring \omp{} support in order
to produce a thread-based parallel programs and discarded by the
others.
\omp{} (standard) features three different classes of mechanisms to
express and manage various aspects of parallelism, respectively to:
\begin{itemize}
\item  identify and distribute parallel computations among
  different resources;
\item manage scope and ownership of the program data;
\item introduce synchronisation to avoid race conditions.
\end{itemize}

Program flows-of-control (e.g. threads, processes and any similar
entity) are not directly managed by language directives.  Programmers
highlight program sections that should be parallelised. The \omp{}
support automatically defines threads and synchronisations in order to
produce an equivalent parallel program implementation. In the scale
introduced in Sec.~\ref{sec:intro}, it exhibits a medium/high abstraction level.

The main \omp{} mechanism is the \code{parallel} pragma, which is used to
bound pieces of sequential code which are going to be computed in
parallel by a team of threads. Inside a \code{parallel} section, in
order to specify a particular parallel paradigm that the team thread
have to implement, specific directives can be inserted.  The
\code{parallel for} directive expresses data parallelism while the
\code{section} functional parallelism. Because of the limitation of the
\code{section} mechanism, which provides only static functional
partition, from the version 3.0 \omp provides a new construct called
\code{task} to model independent units of work which are automatically
scheduled without programmers intervention. As suggested in
\cite{openmpTask} \code{task} construct can be used to build a web
server, and since web servers exhibits a typical streaming
behaviour, we will use the \code{task} construct to build our farm schema.

\subsubsection{Pseudo-code}
\omp{} do not natively include a farm skeleton, which should be
realised using lower-level features, such as the \code{task}
construct.  Our \omp{} farm schema is shown
in Fig.~\ref{code:farm_omp}. The schema is quite simple; a \code{single} section is
exploited to highlight the Emitter behaviour. The new independent
tasks, defined by the Emitter, are marked with the \code{task}
directive in order to leave their computation scheduling to the \omp
run time support.

The Collector is implemented in a different way. Instead of
implementing it with a \code{single} section (as 
for the Emitter), and therefore introducing an explicit locking mechanism for
synchronisation between workers and Collector, we realise Collector
functionality by means of workers cooperative behaviour: they simply
output tasks using an \omp \code{critic} section. 
This mechanism enable us to output tasks from the stream without
introducing any global synchronisation (barrier).

\lstnewenvironment{openMP}[2] 
{\lstset{ %
language=C++,                % choose the language of the code
basicstyle=\ttfamily\footnotesize,       % the size of the fonts that are used for the code
%numbers=left,                   % where to put the line-numbers
%numberstyle=\footnotesize,      % the size of the fonts that are used for the line-numbers
%stepnumber=1,                   % the step between two line-numbers. If it's 1 each line will be numbered
%numbersep=0.5pt,                  % how far the line-numbers are from the code
backgroundcolor=\color{white},  % choose the background color. You must add \usepackage{color}
showspaces=false,               % show spaces adding particular underscores
showstringspaces=false,         % underline spaces within strings
showtabs=false,                 % show tabs within strings adding particular underscores
%frame=single,                   % adds a frame around the code
frame=shadowbox
tabsize=2,                      % sets default tabsize to 2 spaces
captionpos=b,                   % sets the caption-position to bottom
breaklines=true,                % sets automatic line breaking
breakatwhitespace=false,        % sets if automatic breaks should only happen at whitespace
escapeinside={\%*}{*)},          % if you want to add a comment within your code
%MIEI =P
columns=flexible,
mathescape=true,
morekeywords={pragma,omp,single,task,task,nowait,untied,critic},
%backgroundcolor=\color{green},
%#1,
%label={code:#2}
}}{}
%[caption={\lc code of the emitter process in $EC$
%    configuration}, label=emitter]
\begin{figure}[t]
\begin{openMP}{}{}
int main (int argc, char *argv[]) 
{
  #pragma omp parallel private($newtask$)
  {
    /* EMITTER */
    #pragma omp single nowait 
    {
      while($\exists\;newtask$){
        $newtask$ = create_task();
        /* WORKER */
        #pragma omp task untied
        {
          compute_task($newtask$);
          /* COLLECTOR */
          #pragma omp critic
          {
            collect_task($newtask$);
          }
        }
      }
    }
  }
\end{openMP}
\label{code:farm_omp}
\caption{Farm structure in \omp.}
\end{figure}

\subsection{Cilk Farm}
\label{cilk}
Cilk is a language for multi-threaded parallel programming that extends
the C language. Cilk provides programmers with mechanisms to
spawn independent flows of controls \emph{(cilk-threads)} according to
the \emph{fork/join} model. The scheduling of the computation of flows is
managed by a efficient work-stealing scheduler \cite{BlumofeJoKu96}.

Cilk controls flows are supported by a share memory featured by a DAG
consistency \cite{BlumofeFrJo96}, which is a quite relaxed consistency
model. Cilk-threads synchronise according to the DAG consistency at
the join (\emph{sync} construct), and optionally, atomically
execute a sort of call-back function (\emph{inlet} procedure).

Cilk lock variables are provided to define atomic chunks of code,
enabling programmers to address synchronisation patterns that cannot
be expressed using DAG consistency. As matter of fact, Cilk lock
variables represent an escape in a programming model which has been
designed for avoiding critical regions.

\subsubsection{Pseudo-code}
Our reference code structure for a farm implemented in Cilk is
introduced in Fig.~\ref{code:cilk}. A thread is spawn at the
beginning of the program to implement the emitter behaviour and remain
active until the end of the computation.  The emitter thread defines
new tasks and spawn new threads for their computation.

To avoid explicit lock mechanism we target \code{inlet} constructs.
Ordinarily, a spawned Cilk thread can return its results only to the
parent thread, putting those results in a variable in the parent's
frame. The alternative is to exploit an \code{inlet}, which is a function
internal to a Cilk procedure to handle the results of a spawned
thread call as it returns. One major reason to use inlets is that
all the inlets of a procedure are guaranteed to operate atomically
with regards to each other and to the parent procedure, thus avoiding
race conditions that can come out when the multiple returning threads try
to update the same variables in the parent frame. 

The \code{inlet}, which can be compared with \omp{} \code{critic}
sections, can be easily exploited to implement the Collector behaviour
as presented in the definition of the emitter function in Fig.~\ref{code:cilk}.

Because \code{inlet} feature the limitation that the function have to be
called from the cilk procedure that hosts the function, our emitter procedure, and our worker procedure have to be the same to use \code{inlet}. 
We differentiate the two behaviour exploiting a tag parameter and switching 
on its value.

\lstnewenvironment{cilk}[2] {\lstset{ %
    language=C++, % choose the language of the code
basicstyle=\ttfamily\footnotesize,  % the size of the fonts that are used for the code
%numbers=left,                   % where to put the line-numbers
%numberstyle=\footnotesize,      % the size of the fonts that are used for the line-numbers
%stepnumber=1,                   % the step between two line-numbers. If it's 1 each line will be numbered
%numbersep=5pt,                  % how far the line-numbers are from the code
backgroundcolor=\color{white},  % choose the background color. You must add \usepackage{color}
showspaces=false,               % show spaces adding particular underscores
showstringspaces=false,         % underline spaces within strings
showtabs=false,                 % show tabs within strings adding particular underscores
%frame=single,                   % adds a frame around the code
frame=shadowbox
tabsize=2,                      % sets default tabsize to 2 spaces
captionpos=b,                   % sets the caption-position to bottom
breaklines=true,                % sets automatic line breaking
breakatwhitespace=false,        % sets if automatic breaks should only happen at whitespace
escapeinside={\%*}{*)},          % if you want to add a comment within your code
%MIEI =P
mathescape=true,
columns=flexible,
morekeywords={spawn,cilk,inlet,main},
%backgroundcolor=\color{green},
%#1,
%label={code:#2}
}}{}
%[caption={\lc code of the emitter process in $EC$
%    configuration}, label=emitter]
\begin{figure}[t]
  \begin{cilk}{}{}
cilk int * emitter(int * $newtask$, int tag) {    
  inlet void collector(int * $newtask$) {	
    collect_task($newtask$);
  }
  switch(tag) {
    case $WORKER$: {
      compute_task($newtask$);
    }break;
    case $EMITTER$: {
      while($\exists\;newtask$){
        $newtask$ = create_task();
        collector(spawn emitter($newtask$, $WORKER$));
      }
    }break;
    default: ;
  }
  return $newtask$;
}

cilk int main(int argc, char *argv[]) {
  null = spawn emitter(NULL, $EMITTER$);
  sync;
}
 \end{cilk}
\label{code:cilk}
\caption{Farm structure in Cilk.}

\end{figure}

\subsection{TBB Farm}
Intel \emph{Threading Building Blocks} (TBB) is a C++ template library
consisting of \emph{containers} and \emph{algorithms} that abstract  the
usage of native threading packages (e.g. POSIX threads) in which
individual threads of execution are created, synchronised, and
terminated manually. Instead the library abstracts access to the
multiple processors by allowing the operations to be treated
as \emph{tasks}, which are allocated to individual cores dynamically
by the library's run-time engine, and by automating efficient use of
the cache. The tasks and synchronisations among them are extracted
from language constructs  such as \code{parallel\_for},
\code{parallel\_reduce}, \code{parallel\_scan}, and \code{pipeline}. Tasks might also
cooperate via shared memory through concurrent containers
(e.g. \code{concurrent\_queue}), several flavours of mutex
(lock, and atomic operations (e.g. Compare\_And\_Swap)
\cite{intel:skeletons:09,tbb:book:2007}.  

This approach groups TBB in a family of solutions for parallel
programming aiming to enable programmers to explicitly define
parallel behaviour via parametric exploitation patterns  (\emph{skeletons},
actually) that have been widely explored the last two decades both
for distributed memory \cite{cole-th, p3l:hp:92,
  lithium:fgcs:03,serot:taggedtoken:ppl:2001,mapreduce:google:04, assist:cunhabook:05} and
shared memory  \cite{eskimo:PPL:03} programming models.

\subsubsection{Pseudo-code}

\lstnewenvironment{TBB}[2] 
{\lstset{ %
language=C++,                % choose the language of the code
basicstyle=\ttfamily\footnotesize,        % the size of the fonts that are used for the code
%numbers=left,                   % where to put the line-numbers
%numberstyle=\footnotesize,      % the size of the fonts that are used for the line-numbers
%stepnumber=1,                   % the step between two line-numbers. If it's 1 each line will be numbered
%numbersep=5pt,                  % how far the line-numbers are from the code
backgroundcolor=\color{white},  % choose the background color. You must add \usepackage{color}
showspaces=false,               % show spaces adding particular underscores
showstringspaces=false,         % underline spaces within strings
showtabs=false,                 % show tabs within strings adding particular underscores
%frame=single,                   % adds a frame around the code
frame=shadowbox
tabsize=2,                      % sets default tabsize to 2 spaces
captionpos=b,                   % sets the caption-position to bottom
breaklines=true,                % sets automatic line breaking
breakatwhitespace=false,        % sets if automatic breaks should only happen at whitespace
escapeinside={\%*}{*)},          % if you want to add a comment within your code
%MIEI =P
mathescape=true,
columns=flexible,
morekeywords={main,tbb, pipeline,tbb::, task_scheduler_init, blocked_range,auto_partitioner, filter,add_filter,parallel_for},
%backgroundcolor=\color{green},
%#1,
%label={code:#2}
}}{}
\begin{figure}[!h]
\begin{TBB}{}{}
class Emitter:public tbb::filter {
public:
   Emitter(const int grain):tbb::filter(tbb::serial_in_order),grain(grain) {}
   void * operator()(void*) {
     $newtask_t$ * Task[grain];
     while($\exists\;newtask$){       
       Task = create_task();
       return Task;
     }
     return NULL; //EOS
   }
};

class Compute {
  task_t ** Task;
public:
  Compute(task_t ** Task):Task(Task){}
  void operator() (const tbb::blocked_range<int>& r) const {
    for (int i=r.begin();i<r.end();++i) 
       compute_task(Task[i]);  
  }
};

class Worker:public tbb::filter {
 task_t ** Task;
public:
   Worker(const int grain):tbb::filter(tbb::serial_in_order),grain(grain) {}
   void * operator()(void * T) {
     Task = static_cast<task_t**>(T);
     tbb::parallel_for(tbb::blocked_range<int>(0,grain),
                       Compute(Task),
                       tbb::auto_partitioner());
     return Task;
   }
};

int main (int argc, char *argv[])  {
  Emitter   E(PIPE_GRAIN);
  Worker    W(PIPE_GRAIN);
  tbb::task_scheduler_init init(NUMTHREADS);
  tbb::pipeline pipeline;
  pipeline.add_filter(E);
  pipeline.add_filter(W);
  pipeline.run(1);
}
\end{TBB}
\caption {Farm structure in TBB. \label {code:farm_tbb}}
\end{figure}

The structure of the farm paradigm using the TBB library is sketched in Fig.~\ref{code:farm_tbb}. The implementation
is based on the \code{pipeline} construct. The pipeline is composed of 
three stages: Emitter, Worker, Collector. The corresponding three objects 
are registered with the \code{pipeline} object in order to instantiate
the correct communication network.
The Emitter stage produces a pointer to arrays of basic tasks, 
referred as \code{Task} in the pseudo-code, each one of length \code{PIPE\_GRAIN} (for our experiments we set the \code{PIPE\_GRAIN} to 1024). 
The Worker stage is actually a filter that allows the execution of the 
\code{parallel\_for} on input tasks. The \code{parallel\_for} is executed 
using the \code{auto\_partitioner} algorithm provided by the TBB library, 
this way the correct splitting of the \code{Task} array in chunks of 
basic tasks which are assigned to the executor threads, is left to the run-time 
support.

\section{Experiments and Results}
\label{sec:exp}
We compare the performance of FastFlow farm implementation against
OpenMP, Cilk and TBB using two families of
applications: a synthetic micro-benchmark and the Smith-Waterman local
sequence alignment algorithm.  All experiments reported in the
following sections are executed on a shared memory Intel platform with
2 quad-core Xeon E5420 Harpertown @2.5GHz with 6MB L2 cache and 8
GBytes of main memory.

\begin{figure}
\begin{center}
\includegraphics[width=0.7\linewidth]{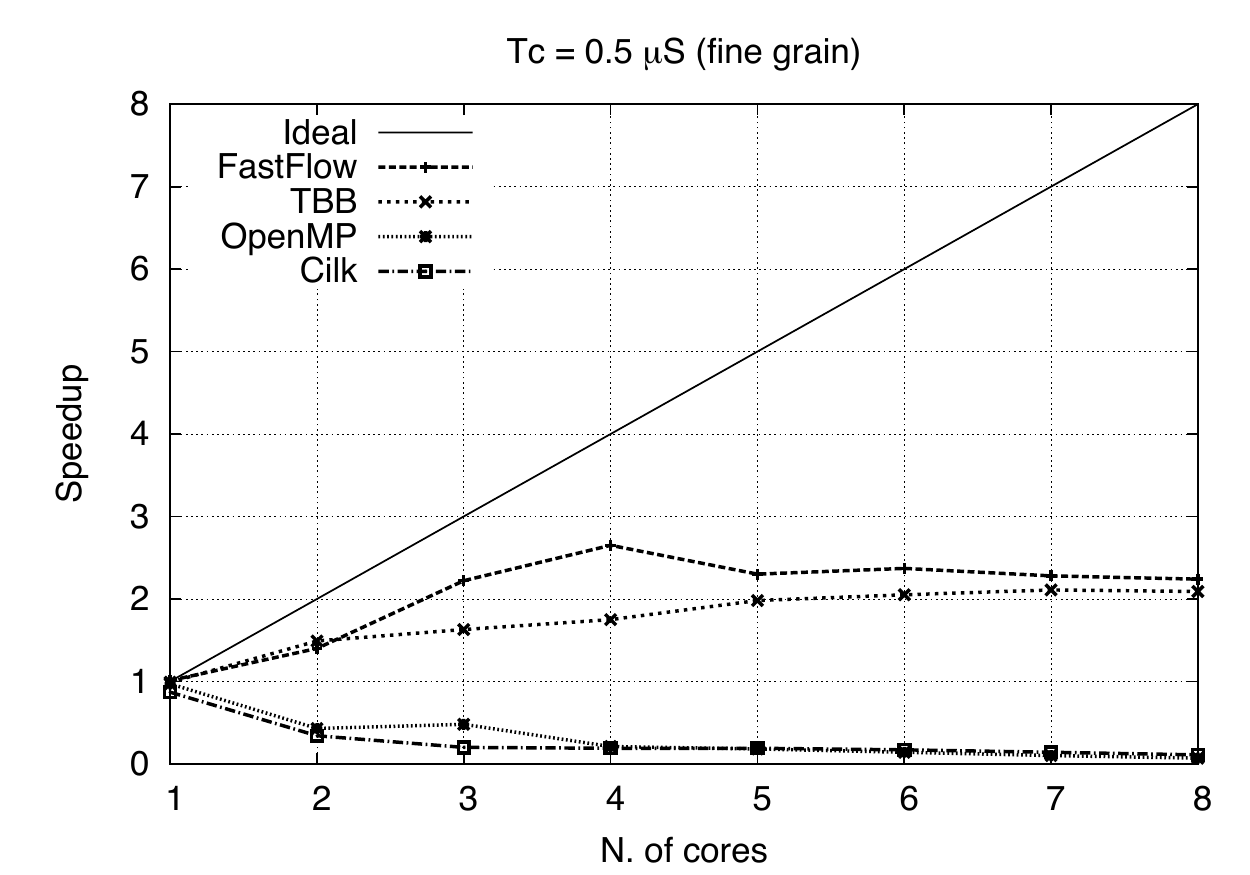}
\includegraphics[width=0.7\linewidth]{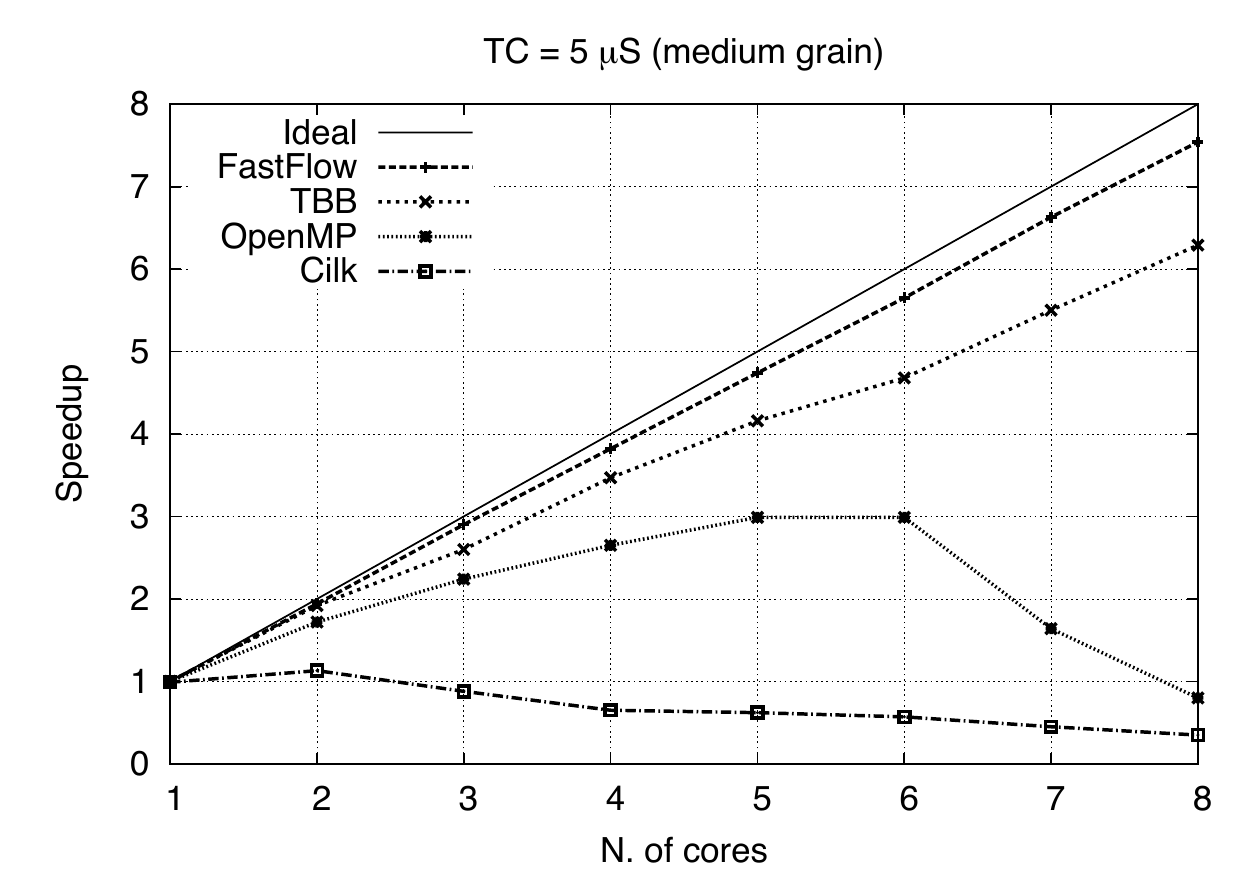}
\includegraphics[width=0.7\linewidth]{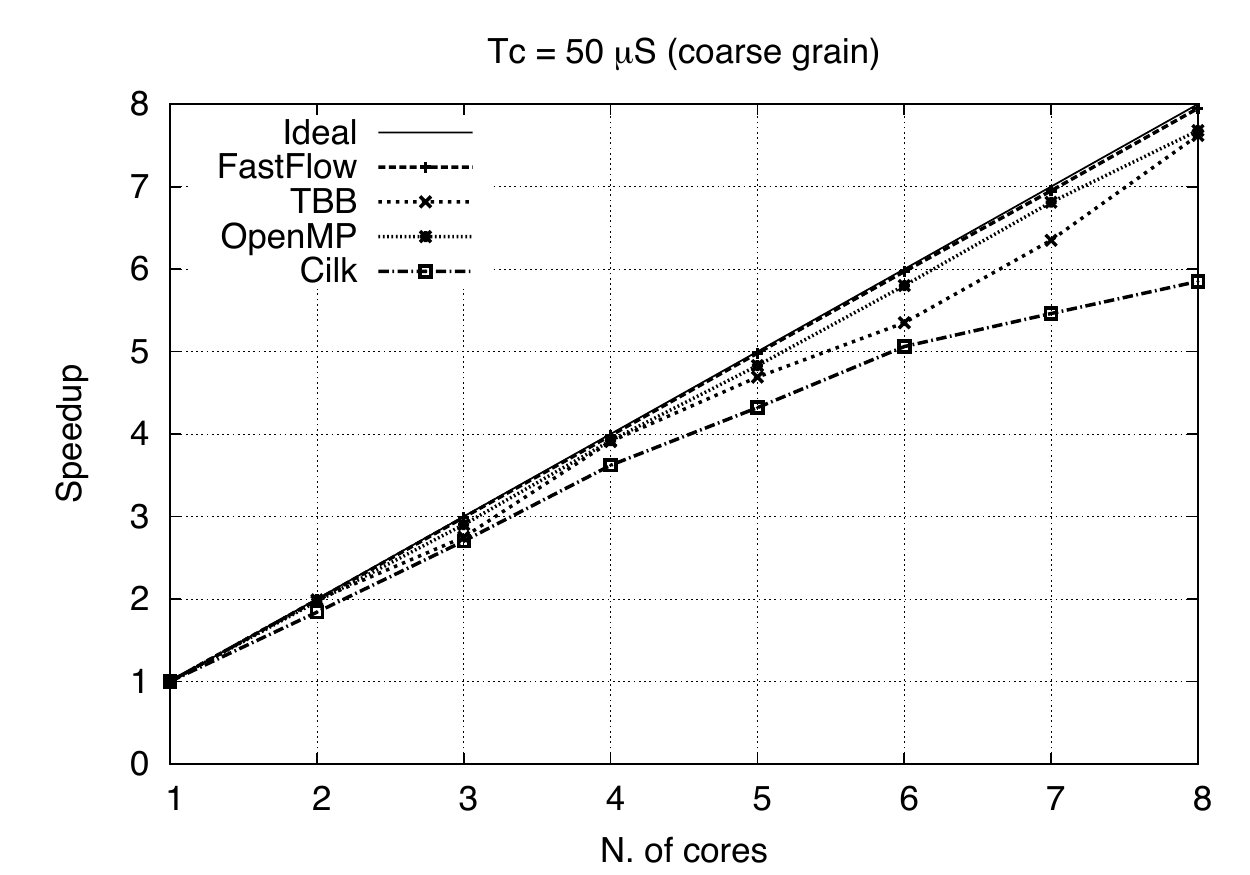}
\caption{The speedup of different implementations of the farm paradigm
  for different computational grains, where $Tc$ is the Computation
  Time per task: FastFlow vs OpenMP vs TBB vs Cilk. \label{fig:speedup}} 
\end{center}
\end{figure}

\subsection{Farm Communication Overhead}

In order to test the overhead of the communication infrastructure for
the different implementations of the farm construct, we developed a
simple micro-benchmark application that emulate a typical parallel
filtering application via a farm skeleton.  The stream is composed by
a ordered sequence of tasks which have a synthetic computational load
associated. Varying this load is possible to evaluate the speedup of
the paradigm for different computation grains.  Each task is
dynamically allocated by the emitter entity and freed by the Collector
one. It consists of an array of 10 memory words that the worker
reads and updates before passing the task to the Collector entity.
Furthermore, each worker spend a fixed amount of time that correspond
to the synthetic workload.

When possible, we have used the best parallel allocator
(i.e. allocator which implementation is optimised to be used in
parallel) available for the dynamic memory allocation. For example in
the case of TBB, we used the TBB
\code{scalable\_allocator}.

In the OpenMP and Cilk cases, where no optimised parallel allocator are
provided, we exploited the standard libc allocator.  Anyway, with
respect to the previously presented micro-benchmark, the performance
degradation associated to an allocator, which is not optimised for parallel
utilisation, is not determining. In fact the maximum parallelism required by
the allocator is two: only one thread performs \code{malloc}
operations (i.e. the emitter) and only one thread that performs
\code{free} ones (i.e. the Collector).

As it is evident from the comparison of the trend of the curves in
Fig.~\ref{fig:speedup}, 
the FastFlow implementation exhibits the
best speedup in all cases.  Is it interesting to notice that, for very
fine grain computations (e.g. 0.5 $\mu$S), the OpenMP and Cilk
implementations feature, with the increasing of the parallelism
degree, a speedup factor lower than one: the addition of more workers
just introduces only overhead, therefore leading to performances that are worst
than the sequential ones.

In streaming computation with this type of computation grain, the
communication overhead between successive stages in the farm construct
is the most important limiting factor. Therefore we can assert that
FastFlow is effective for streaming networks because it has been
developed and implemented in order to provide communications with
extremely low overhead as proved by the collected results.

\subsection{Smith-Waterman Algorithm}

In bioinformatics, sequence database searches are used to find the
similarity between a query sequence and subject sequences in the
database in order to determining similar regions between two nucleotide
or protein sequences, encoded as a string of characters.
The sequence similarities can be determined by computing their optimal local
alignments using the Smith-Waterman (SW) algorithm \cite{Waterman81}. 
SW is a dynamic programming algorithm that guaranteed to find the optimal local alignment 
with respect to the scoring system being used. Instead of looking 
at the total sequence, it compares segments of all
possible lengths and optimises the similarity measure.
The costs of this approach is expensive in terms of computing time 
and memory space used due to the rapid growth of biological sequence 
databases (the UniProtKB/Swiss-Prot database Release 57.5 of 07-Jul-09 
contains 471472 sequence entries, comprising 167326533 amino acids) \cite{protein:db:web}. 

The recent emergence of multi- and many-core architectures provides 
the opportunity to significantly reduce the computation time for many
costly algorithms like the Smith-Waterman one.
Recent works in this area focus on the implementation of the SW algorithm
on many-core architectures like GPUs \cite{cudasw} and Cell/BE \cite{FarrarCell} 
and on multi-core architectures exploiting the SSE2 instruction set \cite{FarrarSSE2, SWPS3}.
Among these implementations, we selected the SWPS3  \cite{SWPS3:08} an
optimised extension of the
Farrar's work presented in \cite{FarrarSSE2} of the Strip Waterman-Algorithm for the Cell/BE and on x86/64  
CPUs with SSE2 instructions.  
The original SWPS3 version is designed as a master-worker computation where the 
master process distribute the workload toward a set of worker processes. 
The master process read the query sequence, initialise the data structures needed for the SSE2 computation,
and then fork all the worker processes so that each worker has its own copy of the data. 
All the sequences in the reference database are read and sent to the worker processes over POSIX pipes. 
The worker computes the alignment score of the query with the database sequence provided by the master process,
and sent back over a pipe the resulting score.

\begin{table}
\begin{center}
\textsf{\scriptsize
\begin{tabular}{crrrr}
\toprule
&&\multicolumn{3}{c}{Stream Task Time}\\
\cmidrule{3-5}
Query& Query len& Min ($\mu$S)& Max ($\mu$S)& Avg ($\mu$S)\\
\midrule
P02232 & 144& 0.333	&  2264.9	&25.0\\
P10635& 497& 0.573	&  15257.6	&108.0\\
P27895& 1000 &	0.645	& 16011.9	&197.0\\
P04775& 2005 &	0.690	& 21837.1	&375.0\\
P04775& 5478 & 3.891	& 117725.0	&938.5\\
\bottomrule
\end{tabular}
}
\end{center}
\caption{Minimum, maximum and average computing time  for a selection of query sequences tested using a penalty gap 5-2 k.}
\label{table:Tc}
\end{table}

\begin{figure*}
\begin{center}
\includegraphics[width=0.9\linewidth]{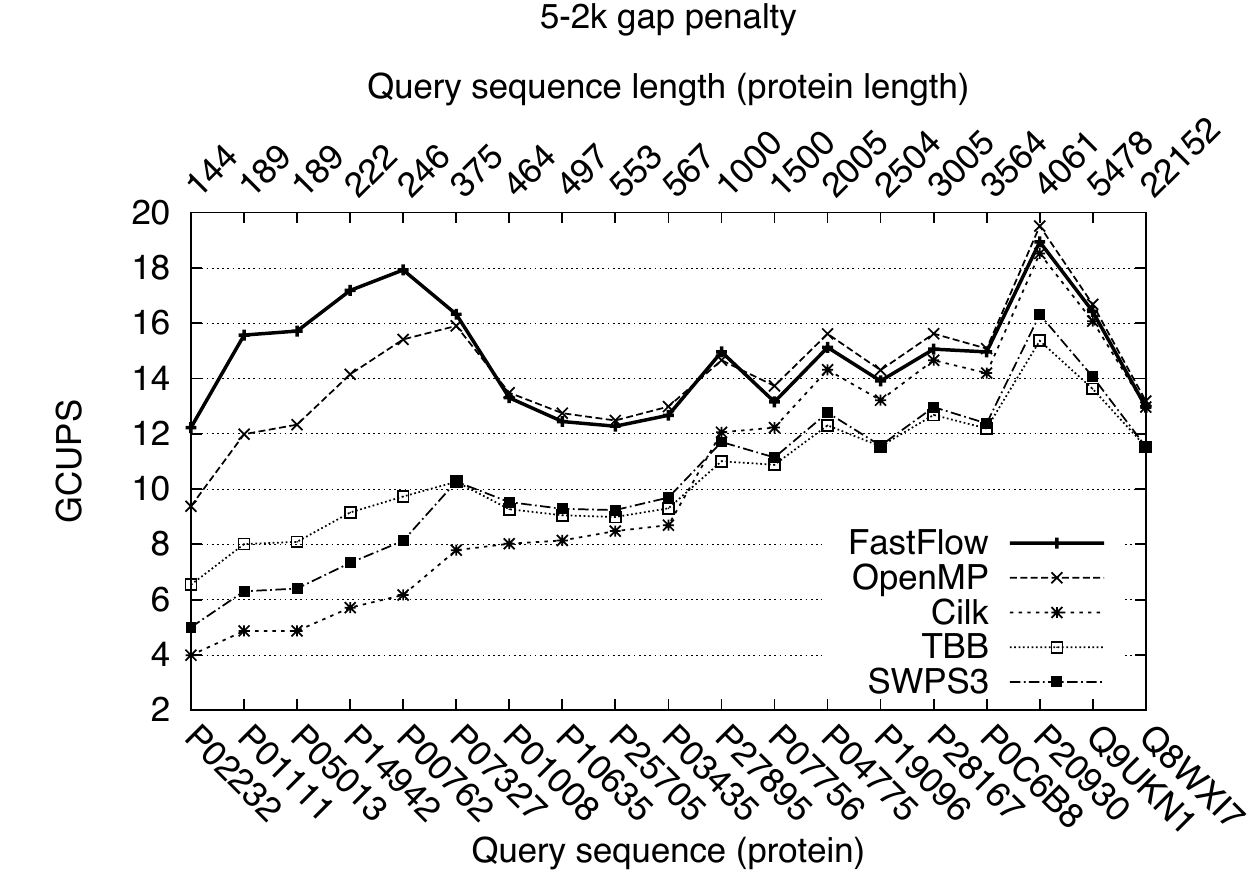}
\includegraphics[width=0.9\linewidth]{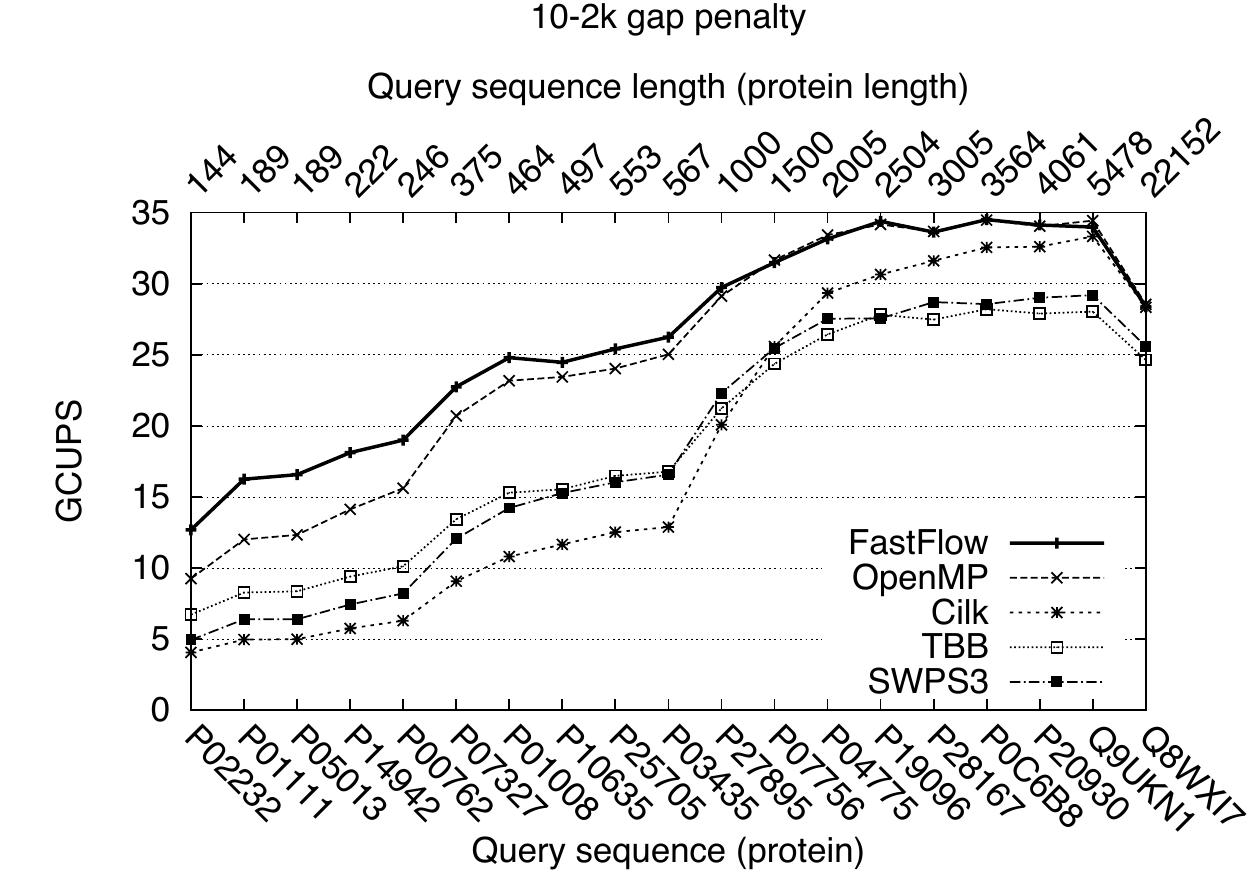}
\caption{Smith-Waterman sequence alignment algorithm: comparison 
between FastFlow, \omp, TBB, and Cilk implementations.
The SWPS3, which is based on POSIX primitives, is the original version
from which the other has been  
derived. All the implementations share exactly the same sequential 
(x86/SSE2 vectorised) code. \label{fig:sw}} 
\end{center}
\end{figure*}

The computational time is sensitive with respect to the query length used for the matching,
the scoring matrix (in our case BLOWSUM50) and the gap penalty. As can be seen from Table 
\ref{table:Tc} very small sequences require a smaller service time with respect
to the longest one. Notice the high variance in the task service time 
reported in the table, this is due to the very different length of the subject sequence 
in the reference database (the average sequence length in UniProtKB/Swiss-Prot is 352 amino acids,
the shortest sequence comprise 2 amino acids whereas the longest one 35213 amino acids).
Furthermore, the higher the gap open and gap extension penalties, the fewer iterations are 
needed for the calculation of the single cell of the similarity score matrix. In our tests
we used the scoring matrix BLOSUM50 with two gap penalty range: 10-2k and 5-2k.

We rewrote the original SWPS3 code in OpenMP, Cilk, TBB and FastFlow
following the schemata presented before. In this, we did not modify
the sequential code at all to achieve a fair comparison.  For
performance reasons, it is important to provide each worker threads
with a copy of the data structures needed for the SSE2
computation. This is a critical aspects especially for implementations
in Cilk and TBB, which do not natively support any kind of
\emph{Thread-Specific-Storage} (TSS). Notwithstanding this data is
read-only, the third-party SSE somehow seems triggering the cache
invalidation accessing the data, which seriously affect the
performance.  To overcome this problem we exploit a tricky solution:
we use TSS exploiting a lower-level with respect to the programming
model.  In OpenMP this is not a problem because we have the
possibility to identify the worker thread with the library call
\code{omp\_get\_thread\_num()}. The same possibility to identify a
thread is offered by FastFlow framework as each parallel entity is
mapped on one thread.

The Emitter entity reads the sequence database and produce a 
stream of pairs: \emph{$\langle$query sequence, subject sequence$\rangle$}. 
The query sequence remains the 
same for all the subject sequences contained in the database. 
The Worker entity computes the striped Smith-Waterman algorithm on the input pairs 
using the SSE2 instructions set. The Collector entity gets the resulting 
score and produce the output string containing the score and the sequence name.

To remove the dependency on the query sequences and the databases used for the 
tests, Cell-Updates-Per-Second (CUPS) is a commonly used performance measure
in bioinformatics. A CUPS represents the time for a complete computation of 
one cell in the matrix of the similarity score, including all memory operations.
Given a query sequence of length Q and a database of size D, the GCUPS (billion Cell Updates 
Per Second) value is calculated by: 
$$GCUPS = \frac{|Q|\ |D|}{T\ 10^9}$$

where T is the total execution time in seconds. 
The performance of the different SW algorithm implementations 
has been benchmarked and analysed by searching for 19 sequences of length from
144 (the P02232 sequence) to 22,142 (the Q8WXI7 sequence) against Swiss-Prot 
release 57.5 database. The tests has been carried out on a dual quad-core Intel 
Xeon @2.50GHz running the Linux OS (kernel 2.6.x). 

Figure~\ref{fig:sw} reports the performance comparison between FastFlow,
OpenMP, Cilk, TBB and SWPS3 version of SW algorithm for x86/SSE2 executed on 
the test platform described above. 

As can be seen from the figures, the FastFlow implementation outperforms the
other implementations for short query sequences. The smallest the 
query sequences are the bigger the performance gain is. This is
mainly due to lower overhead of FastFlow communication channels with respect to the other implementations; short sequences require a smaller service time.

Cilk obtains lower performance value with respect to the original SWPS3 version with small sequences while performs very well with longer ones.
OpenMP offers the best performance after FastFlow. Quite surprisingly 
TBB does not obtain the same good speedup that has been obtained with the 
micro-benchmark. It is still not clear which are the reasons, further investigation is required to find out the overhead source in the TBB version.

\section{Conclusions}

In this work we have introduced FastFlow, a low-level template library
based on lock-free communication channel explicitly designed
to support low-overhead high-throughput streaming applications on
commodity cache-coherent multi-core architectures.
We have shown that FastFlow can be directly used to implement complex 
streaming applications exhibiting cutting-edge performance on
a commodity multi-core. 

Also, we have demonstrated that FastFlow makes
it possible the efficient parallelisation of third-party legacy code, as the
x86/SSE vectorised Smith-Waterman code.
In the short term, we envision FastFlow as middleware tier of a
``skeletal'' high-level programming framework that will discipline the
usage of efficient network patterns, possibly extending an existing
programming framework (e.g. TBB) with stream-specific constructs.
As this end, we studied how  a streaming farm can be realised
using several state-of-the-art programming frameworks for
multi-core, and we have experimentally demonstrated that FastFlow farm is
faster than other farm implementations on both synthetic benchmark and
Smith-Waterman application. 

As expected,  the performance edge of
FastFlow over the other frameworks is bold for fine-grained
computations. This makes FastFlow suitable to implement a fast
macro data-flow executor (actually wrapping around the order
preserving farm), thus to achieve the automatic
parallelisation of many classes of algorithms, including dynamic
programming \cite{vigoni:fut_rmi:book:05}.
FastFlow will be released as open source library.

A preliminary version of this work has been presented at the ParCo
conference \cite{fastflow:parco:09}.

\section{Acknowledgments}

We thank Marco Danelutto, Marco Vanneschi and Peter Kilpatrick for the
many insightful discussions.

This work was partially funded by the project BioBITs (``Developing
White and Green Biotechnologies by Converging Platforms from Biology
and Information Technology towards Metagenomic'') of Regione Piemonte,
by the project FRINP of the ``Fondazione della Cassa di Risparmio di
Pisa'', and by the WG Ercim CoreGrid topic ``Advanced Programming
Models''.

%\bibliographystyle{plain}
%\bibliography{UniPisaGroup,grid,ac,sw,multicore}

\begin{thebibliography}{10}

\bibitem{eskimo:PPL:03}
Marco Aldinucci.
\newblock eskimo: experimenting with skeletons in the shared address model.
\newblock {\em Parallel Processing Letters}, 13(3):449--460, September 2003.

\bibitem{assist:imp:europar:03}
Marco Aldinucci, Sonia Campa, Pierpaolo Ciullo, Massimo Coppola, Silvia Magini,
  Paolo Pesciullesi, Laura Potiti, Roberto Ravazzolo, Massimo Torquati, Marco
  Vanneschi, and Corrado Zoccolo.
\newblock The implementation of {ASSIST}, an environment for parallel and
  distributed programming.
\newblock In {\em Proc. of 9th Intl Euro-Par 2003 Parallel Processing}, volume
  2790 of {\em {LNCS}}, pages 712--721, Klagenfurt, Austria, August 2003.
  {Springer}.

\bibitem{assist:cunhabook:05}
Marco Aldinucci, Massimo Coppola, Marco Danelutto, Marco Vanneschi, and Corrado
  Zoccolo.
\newblock {ASSIST} as a research framework for high-performance grid
  programming environments.
\newblock In {\em Grid Computing: Software environments and Tools}, chapter~10,
  pages 230--256. {Springer}, January 2006.

\bibitem{pdcs:nf:99}
Marco Aldinucci and Marco Danelutto.
\newblock Stream parallel skeleton optimization.
\newblock In {\em Proc. of PDCS: Intl. Conference on Parallel and Distributed
  Computing and Systems}, pages 955--962, Cambridge, Massachusetts, USA,
  November 1999. IASTED, ACTA press.

\bibitem{lithium:sem:CLSS}
Marco Aldinucci and Marco Danelutto.
\newblock Skeleton based parallel programming: functional and parallel semantic
  in a single shot.
\newblock {\em Computer Languages, Systems and Structures}, 33(3-4):179--192,
  October 2007.

\bibitem{vigoni:fut_rmi:book:05}
Marco Aldinucci, Marco Danelutto, Jan D{\"u}nnweber, and Sergei Gorlatch.
\newblock Optimization techniques for skeletons on grid.
\newblock In {\em Grid Computing and New Frontiers of High Performance
  Processing}, volume~14 of {\em Advances in Parallel Computing}, chapter~2,
  pages 255--273. {Elsevier}, October 2005.

\bibitem{beske:pdp:09}
Marco Aldinucci, Marco Danelutto, and Peter Kilpatrick.
\newblock Towards hierarchical management of autonomic components: a case
  study.
\newblock In {\em Proc. of Intl. Euromicro PDP 2009: Parallel Distributed and
  network-based Processing}, pages 3--10, Weimar, Germany, February 2009. IEEE.

\bibitem{fastflow:parco:09}
Marco Aldinucci, Marco Danelutto, Massimiliano Meneghin, Peter Kilpatrick, and
  Massimo Torquati.
\newblock {Fastflow}: Fast macro data flow execution on multi-core.
\newblock In {\em Intl. Parallel Computing (PARCO)}, Lyon, France,
  September 2009.

\bibitem{lithium:fgcs:03}
Marco Aldinucci, Marco Danelutto, and Paolo Teti.
\newblock An advanced environment supporting structured parallel programming in
  {Java}.
\newblock {\em Future Generation Computer Systems}, 19(5):611--626, July 2003.

\bibitem{tagged-token:mit:90}
K.~Arvind and Rishiyur~S. Nikhil.
\newblock Executing a program on the mit tagged-token dataflow architecture.
\newblock {\em IEEE Trans. Comput.}, 39(3):300--318, 1990.

\bibitem{openmpTask}
Eduard Ayguad\'{e}, Nawal Copty, Alejandro Duran, Jay Hoeflinger, Yuan Lin,
  Federico Massaioli, Xavier Teruel, Priya Unnikrishnan, and Guansong Zhang.
\newblock The design of openmp tasks.
\newblock {\em IEEE Trans. Parallel Distrib. Syst.}, 20(3):404--418, 2009.

\bibitem{DBLP:journals/ppl/BischofGL03}
Holger Bischof, Sergei Gorlatch, and Roman Leshchinskiy.
\newblock {DatTel}: A data-parallel {C++} template library.
\newblock {\em Parallel Processing Letters}, 13(3):461--472, 2003.

\bibitem{BlumofeFrJo96}
Robert~D. Blumofe, Matteo Frigo, Christopher~F. Joerg, Charles~E. Leiserson,
  and Keith~H. Randall.
\newblock Dag-consistent distributed shared memory.
\newblock In {\em Proc. of the 10th Intl. Parallel Processing Symposium}, pages
  132--141, Honolulu, Hawaii, USA, April 1996.

\bibitem{BlumofeJoKu96}
Robert~D. Blumofe, Christopher~F. Joerg, Bradley~C. Kuszmaul, Charles~E.
  Leiserson, Keith~H. Randall, and Yuli Zhou.
\newblock {C}ilk: An efficient multithreaded runtime system.
\newblock {\em Journal of Parallel and Distributed Computing}, 37(1):55--69,
  August 1996.

\bibitem{Brook}
Ian Buck, Tim Foley, Daniel Horn, Jeremy Sugerman, Kayvon Fatahalian, Mike
  Houston, and Pat Hanrahan.
\newblock Brook for gpus: stream computing on graphics hardware.
\newblock In {\em ACM SIGGRAPH '04 Papers}, pages 777--786, New York, NY, USA,
  2004. ACM Press.

\bibitem{trasnmem:scicomp:05}
Christopher Cole and Maurice Herlihy.
\newblock Snapshots and software transactional memory.
\newblock {\em Sci. Comput. Program.}, 58(3):310--324, 2005.

\bibitem{cole-th}
Murray Cole.
\newblock {\em Algorithmic Skeletons: Structured Management of Parallel
  Computations}.
\newblock Research Monographs in Parallel and Distributed Computing. Pitman,
  1989.

\bibitem{ske:homepage}
Murray Cole.
\newblock {\em Skeletal Parallelism home page}, 2009 (last accessed).
\newblock \url{http://homepages.inf.ed.ac.uk/mic/Skeletons/}.

\bibitem{p3l:hp:92}
Marco Danelutto, Roberto~Di Meglio, Salvatore Orlando, Susanna Pelagatti, and
  Marco Vanneschi.
\newblock A methodology for the development and the support of massively
  parallel programs.
\newblock {\em Future Generation Compututer Systems}, 8(1-3):205--220, 1992.

\bibitem{darlington:parle:93}
J.~Darlington, A.~J. Field, P.G. Harrison, P.~H.~J. Kelly, D.~W.~N. Sharp,
  R.~L. While, and Q.~Wu.
\newblock Parallel programming using skeleton functions.
\newblock In {\em Proc. of Parallel Architectures and Langauges Europe
  (PARLE'93)}, volume 694 of {\em LNCS}, pages 146--160, Munich, Germany, June
  1993. Springer.

\bibitem{mapreduce:google:04}
Jeffrey Dean and Sanjay Ghemawat.
\newblock {MapReduce}: Simplified data processing on large clusters.
\newblock In {\em Usenix OSDI '04}, pages 137--150, December 2004.

\bibitem{FarrarCell}
Michael Farrar.
\newblock {Smith-Waterman} for the cell broadband engine.

\bibitem{FarrarSSE2}
Michael Farrar.
\newblock Striped {Smith-Waterman} speeds database searches six times over
  other simd implementations.
\newblock {\em Bioinformatics}, 23(2):156--161, 2007.

\bibitem{fastforward:ppopp:08}
John Giacomoni, Tipp Moseley, and Manish Vachharajani.
\newblock Fastforward for efficient pipeline parallelism: a cache-optimized
  concurrent lock-free queue.
\newblock In {\em Proc. of the 13th ACM SIGPLAN Symposium on Principles and
  practice of parallel programming (PPoPP)}, pages 43--52, New York, NY, USA,
  2008. ACM.

\bibitem{intel:skeletons:tbb}
Intel Corp.
\newblock {\em Threading Building Blocks}, 2009.
\newblock \url{http://www.threadingbuildingblocks.org/}.

\bibitem{intel:skeletons:09}
Intel Corp.
\newblock {\em Intel Threading Building Blocks}, July 2009 (last accessed).
\newblock \url{http://software.intel.com/en-us/intel-tbb/}.

\bibitem{CUDA}
David Kirk.
\newblock Nvidia cuda software and gpu parallel computing architecture.
\newblock In {\em Proc. of the 6th Intl. symposium on Memory management (ISM)},
  pages 103--104, New York, NY, USA, 2007. ACM.

\bibitem{mpmc1}
Edya Ladan-mozes and Nir Shavit.
\newblock An optimistic approach to lock-free fifo queues.
\newblock In {\em In Proc. of the 18th Intl. Symposium on Distributed
  Computing, LNCS 3274}, pages 117--131. Springer, 2004.

\bibitem{Lamport}
Leslie Lamport.
\newblock Specifying concurrent program modules.
\newblock {\em ACM Trans. Program. Lang. Syst.}, 5(2):190--222, 1983.

\bibitem{cudasw}
Yongchao Liu, Douglas Maskell, and Bertil Schmidt.
\newblock {CUDASW++}: optimizing {Smith-Waterman} sequence database searches
  for {CUDA}-enabled graphics processing units.
\newblock {\em BMC Research Notes}, 2(1):73, 2009.

\bibitem{mpmc2}
H.~Massalin and C.~Pu.
\newblock Threads and input/output in the synthesis kernal.
\newblock {\em SIGOPS Oper. Syst. Rev.}, 23(5):191--201, 1989.

\bibitem{ABA:98}
Maged~M. Michael and Michael~L. Scott.
\newblock Nonblocking algorithms and preemption-safe locking on multiprogrammed
  shared memory multiprocessors.
\newblock {\em Journal of Parallel and Distributed Computing}, 51(1):1--26,
  1998.

\bibitem{openMP}
Insung Park, Michael~J. Voss, Seon~Wook Kim, and Rudolf Eigenmann.
\newblock Parallel programming environment for openmp.
\newblock {\em Scientific Programming}, 9:143--161, 2001.

\bibitem{kuchen-farm}
M.~Poldner and H.~Kuchen.
\newblock Scalable farms.
\newblock In {\em Proc. of Intl. PARCO 2005: Parallel Computing}, Malaga,
  Spain, September 2005.

\bibitem{idiom:sc:95}
Bill Pottenger and Rudolf Eigenmann.
\newblock Idiom recognition in the {P}olaris parallelizing compiler.
\newblock In {\em Proc. of the 9th Intl. Conference on Supercomputing (ICS
  '95)}, pages 444--448, New York, NY, USA, 1995. ACM Press.

\bibitem{mpmc3}
S.~Prakash, Yann~Hang Lee, and T.~Johnson.
\newblock A nonblocking algorithm for shared queues using compare-and-swap.
\newblock {\em IEEE Trans. Comput.}, 43(5):548--559, 1994.

\bibitem{tbb:book:2007}
James Reinders.
\newblock {\em Intel Threading Building Blocks: Outfitting C++ for Multi-core
  Processor Parallelism}.
\newblock O'Reilly, 2007.

\bibitem{serot:taggedtoken:ppl:2001}
J.~Serot.
\newblock {Tagged-token data-flow for skeletons}.
\newblock {\em Parallel Processing Letters}, 11(4):377--392, 2001.

\bibitem{SWPS3}
Adam Szalkowski, Christian Ledergerber, Philipp Kraehenbuehl, and Christophe
  Dessimoz.
\newblock {SWPS3} - fast multi-threaded vectorized {Smith-Waterman} for {IBM
  Cell/B.E. and x86/SSE2}.
\newblock {\em BMC Research Notes}, 1(1), 2008.

\bibitem{SWPS3:08}
Adam Szalkowski, Christian Ledergerber, Philipp Kr{\"a}henb{\"u}hl, and
  Christophe Dessimoz.
\newblock {\em SWPS3 -- fast multi-threaded vectorized Smith-Waterman for IBM
  Cell/B.E. and x86/SSE2}, 2008.

\bibitem{streamIt}
William Thies, Michal Karczmarek, and Saman~P. Amarasinghe.
\newblock {StreamIt}: A language for streaming applications.
\newblock In {\em Proc. of the 11th Intl. Conference on Compiler Construction
  (CC)}, pages 179--196, London, UK, 2002. Springer-Verlag.

\bibitem{mpmc4}
Philippas Tsigas and Yi~Zhang.
\newblock A simple, fast and scalable non-blocking concurrent fifo queue for
  shared memory multiprocessor systems.
\newblock In {\em SPAA '01: Proc. of the 13th ACM symposium on Parallel
  algorithms and architectures}, pages 134--143, New York, NY, USA, 2001. ACM.

\bibitem{protein:db:web}
UniProt Consortium.
\newblock {\em UniProt web site}, July 2009 (last accessed).

\bibitem{Waterman81}
M.~S. Waterman and T.~F. Smith.
\newblock Identification of common molecular subsequences.
\newblock {\em J. Mol. Biol.}, 147:195--197, 1981.

\end{thebibliography}

\end{document}